\documentclass[aps,prl,twocolumn,showpacs,groupedaddress]{revtex4-1} 

\usepackage{graphicx}
\usepackage{dcolumn} 
\usepackage{bm}      
\usepackage{soul}    
\usepackage{pdfpages}
\usepackage[normalem]{ulem}

\begin{document}

\title{Breakdown of Fermi liquid description for strongly interacting fermions}
\author
{Yoav Sagi$^{\ast}$, Tara E. Drake$^{\ast}$, Rabin Paudel, Roman Chapurin \& Deborah S. Jin$^{\ast}$$^{\ast}$\\
\normalsize{$^{\ast}$These authors contributed equally to this work.}\\
\normalsize{$^\ast$$^\ast$To whom correspondence should be addressed; E-mail:  jin@jilau1.colorado.edu.}
}
\affiliation{JILA, National Institute of Standards and Technology and the University of Colorado, and the Department of Physics, University of Colorado, Boulder, CO 80309-0440, USA}
\date{January 19, 2015}

\begin{abstract}
The nature of the normal state of an ultracold Fermi gas in the BCS-BEC crossover regime is an intriguing and controversial topic. While the many-body ground state remains a condensate of paired fermions, the normal state must evolve from a Fermi liquid to a Bose gas of molecules as a function of the interaction strength. How this occurs is still largely unknown. We explore this question with measurements of the distribution of single-particle energies and momenta in a nearly homogeneous gas above $T_c$. The data fit well to a function that includes a narrow, positively dispersing peak that corresponds to quasiparticles and an ``incoherent background" that can accommodate broad, asymmetric line shapes. We find that the quasiparticle's spectral weight vanishes abruptly as the strength of interactions is modified, which signals the breakdown of a Fermi liquid description. Such a sharp feature is surprising in a crossover.     
\end{abstract}

\maketitle
Landau's Fermi liquid theory is a well-established and powerful paradigm for describing systems of interacting fermions \cite{LandausFermiLiquid1956,LandauLifshitzStatisticalPart2}. It postulates that even in the presence of strong interactions, the system retains a Fermi surface and has low energy excitations that are long-lived, fermionic, and nearly non-interacting. The effect of interactions is incorporated into renormalized properties of these quasiparticle excitations, such as an effective mass, $m^*$, that is larger than the bare fermion mass, $m$, and a spectral weight, or quasiparticle residue, that is between zero and one \cite{LandauLifshitzStatisticalPart2}. While Fermi liquid theory is extremely successful in describing a wide range of materials, 
it fails in systems exhibiting strong fluctuations or spatial correlations. Understanding the origin of such breakdowns of a Fermi liquid description is an outstanding challenge in strongly correlated electron physics \cite{Hill2001}.

An ultracold Fermi gas with tunable interactions is a paradigmatic strongly correlated system. These atomic gases provide access to the crossover from Bardeen-Cooper-Schrieffer (BCS) superconductivity to Bose-Einstein condensation (BEC) of tightly bound fermion pairs \cite{Regal2006,Ketterle2008,Chen20051,BCS-BEC_Zwerger_book,doi:10.1146/annurev-conmatphys-031113-133829}. The question of whether the Fermi liquid paradigm breaks down in the normal state of the crossover is related to the prediction of a ``pseudogap'' phase, where incoherent many-body pairing occurs above the transition temperature $T_c$. This pseudogap phase has bosonic (pair) excitations, in contrast to the fermionic excitations of a conventional Fermi liquid.
In experiments that probed the strongly interacting gas in the middle of the crossover, Fermi-liquid-like behavior was observed in thermodynamic \cite{ISI:000275108400031,Navon07052010} and spin transport properties \cite{ISI:000289469100038}. Meanwhile, evidence for pairing above $T_c$ was reported in photoemission spectroscopy (PES) measurements \cite{Gaebler2010}, which reveal the distribution of single-particle energies and momenta in a many-body system \cite{Stewart2008,RepProgPhys.72.122501}. Interpretation of these data has been controversial, with a Fermi liquid theory and a pseudogap theory each separately argued to agree with the observations \cite{PhysRevLett.106.215303,PhysRevLett.106.060402,PhysRevA.83.061601}. Issues raised include the fact that the PES measurements probed a trapped gas, where averaging over the inhomogeneous density can obscure the intrinsic physics \cite{PhysRevLett.106.215303,PhysRevA.80.063612}, and that thermodynamics measurements are relatively insensitive to a pseudogap compared to spectroscopy. Thus, the question of how a Fermi liquid evolves into a Bose gas of paired fermions in the BCS-BEC crossover, and whether a Fermi liquid description breaks down, remains open. Here, we answer this question with the first PES of a nearly homogeneous Fermi gas; we perform measurements above $T_c$ for a range of interaction strengths through the crossover (Fig. 1a), and find that quasiparticle excitations, which exist on the BCS side, vanish abruptly beyond a certain interaction strength on the BEC side.

\begin{figure*}
	\makebox[\textwidth][c]{\includegraphics[width=16cm]{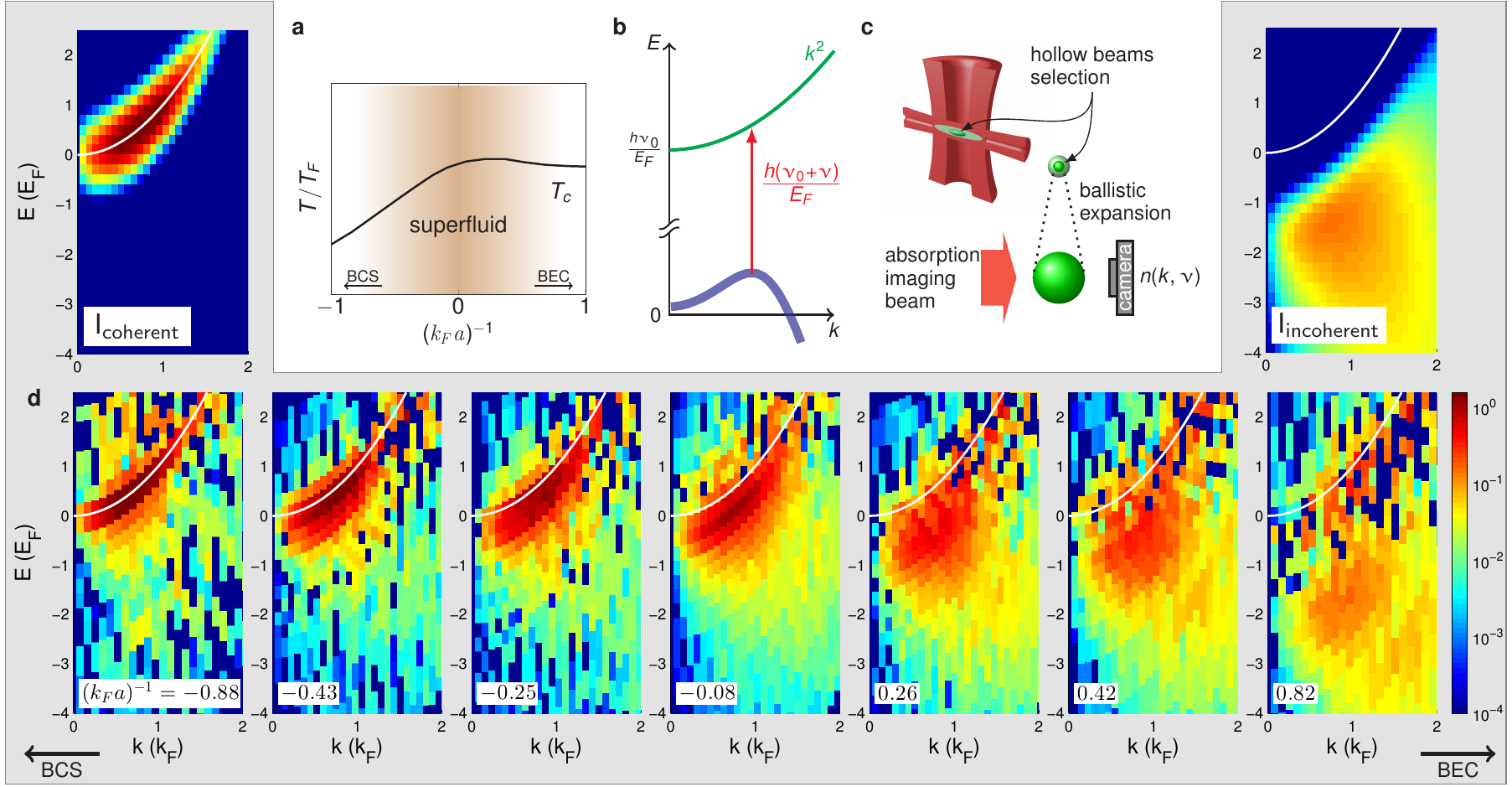}}
	   \caption{   \textbf{Atom PES
	      data.} \textbf{a}, We take data above $T_c$ in the strongly interacting region of the BCS-BEC crossover \cite{PhysRevLett.71.3202}. After initially preparing the gas at small, negative $a$, the magnetic field is swept adiabatically to a final value within the BCS-BEC crossover, where $|(k_Fa)^{-1}|<1$. \textbf{b}, Schematically, an rf photon, which has a negligible momentum, transfers an atom from the strongly interacting state (blue line) to a weakly interacting state (green line). The energy and momentum of the atom in the strongly interacting state are extracted from the measured momentum of the spin-flipped atom and the rf detuning $\nu$ \cite{Stewart2008}. The detuning, $\nu$, of the rf frequency is varied to obtain data for a wide range of $E$ and $k$. \textbf{c}, Immediately following the rf pulse and before time-of-flight expansion, two orthogonally propagating hollow-core beams optically pump atoms at the edges of the outcoupled atom cloud into a dark state \cite{PhysRevA.86.031601,PhysRevLett.109.220402}. The durations of both the rf pulse and optical pumping are short compared to motion of atoms in the trap.
	     \textbf{d (lower panel)}, In these example plots of PES data, the color represents the probability distribution of atoms at a given $E$ and $k$ in the strongly interacting gas. We estimate that the error bar of $(k_F a)^{-1}$ is 0.03. $E=0$ is the energy of a free atom at rest and the white line shows the free-particle dispersion $E=k^2$.	      \textbf{d (upper panel)}, Our two-mode fit function includes a fermionic quasiparticle part, shown on the left with $m^*=1.05$, $E_0=-0.1$, $\mu=0.5$, and $T=0.25$, and an incoherent part, shown on the right with $E_p=1.5$ and $T_p=0.7$.}
    \label{figure1}
\end{figure*}

We prepare a gas of $^{40}\rm{K}$ atoms in an equal mixture of two spin states, where the scattering length, $a$, that parametrizes the interactions is varied using a Fano-Feshbach scattering resonance \cite{RevModPhys.82.1225} (see Fig. 1a and Supplementary Material~\cite{PRL_SOM}).\nocite{PhysRevLett.92.040403,PhDCindyRegal,PhysRevLett.103.210403,PhysRevA.85.012701,PhysRevA.74.033623-2006,PhysRevA.78.053606} To eliminate the complications arising from density inhomogeneity, we combine momentum-resolved rf spectroscopy \cite{Stewart2008} (Fig. 1b) with spatially selective imaging that probes only atoms from the trap center where the density is the highest and has the smallest spatial gradients \cite{PhysRevA.86.031601} (Fig. 1c). The lower panel of Fig. 1d shows PES data taken above $T_c$ for several values of $(k_F a)^{-1}$, where $k_F$ is the Fermi wave number. The PES signal, $I(k,E)$, is proportional to $k^2 A(k,E) f(E)$, where $A(k,E)$ is the atomic spectral function \cite{Stewart2008,PhysRevA.78.033614} and $f(E)$ is the Fermi function. Here, $E$ and $k$ are in units of $E_F$ and $k_F$, respectively, and we normalize each data set so that the integral over all $k$ and $E$ equals 1.

The data in Fig. 1d show an evolution from a positively dispersing, quasiparticle-like spectrum to a broad, negatively dispersing spectrum. Previous trap-averaged atom PES data showed back-bending and large energy widths \cite{Gaebler2010,PhysRevLett.106.060402}. These features are also apparent in the nearly homogeneous data \cite{PRL_SOM}. However, these data are more amenable to quantitative analysis because $k_F$ (and $E_F$) are approximately single-valued across the sample. Similar to the analysis done in electron systems, we use a two-mode function to describe the PES signal \cite{RevModPhys.75.473}:
\begin{equation}
	I(k,E)=Z I_{\rm{coherent}}(k,E)+(1-Z) I_{\rm{incoherent}}(k,E) \ \ ,
	\label{fitting_function}
\end{equation}
where the first part describes quasiparticles with a positive dispersion, the second part accommodates an ``incoherent background'' that exhibits negative dispersion, and $Z$ is the quasiparticle spectral weight. When these two parts (defined below) are combined, the resulting dispersion can exhibit back-bending.

The quasiparticles in Fermi liquid theory are long-lived and therefore give rise to narrow energy peaks, which, in principle, could be directly observed. However, such peaks would be broadened by our experimental resolution of about $0.25 E_F$. This resolution is set by the number of atoms (with $E_F$ scaling only weakly with increasing $N$) and the rf pulse duration (see Supplementary Material~\cite{PRL_SOM}), which must be short compared to the harmonic trap period in order to probe momentum states. We convolve Eqn. \ref{fitting_function} with a Gaussian function that accounts for our energy resolution before fitting to the data in order to determine the spectral weight of the quasiparticles (Fig. 1d, upper panel).

To describe quasiparticles, we use
\begin{widetext}
\begin{equation}
I_{coherent}(k,E)=4\pi k^2\cdot\delta(E-\frac{k^2}{m^*}-E_0)\frac{\left[-\left(\pi m^*T\right)^{3/2}\mathrm{Li}_{3/2}\left(-\exp\left(\frac{-E_0+\mu}{T}\right)\right)\right]^{-1}}{\exp\left({\frac{E-\mu}{T}}\right)+1} \ \ ,\end{equation} 
\end{widetext}
which consists of a quadratic dispersion of sharp quasiparticles multiplied by a normalized Fermi distribution ($\delta$ is the Dirac delta function, and $\mathrm{Li}$ is the polylogarithm function). We include as fit parameters, a Hartree shift $E_0$, effective mass $m^*$, chemical potential $\mu$, and temperature $T$. Here, energies are given in units of $E_F$ and $m^*$ in units of $m$. 
This description of Fermi liquid quasiparticles is typically only used very near $k_F$ and for $T$ approaching zero, whereas we fit to data for a larger range in $k$ and with temperatures near 0.2 $T_F$ (just above $T_c$). The latter is necessitated by
the unusually large interaction energy compared to $E_F$, and we note that 
0.2 $T_F$ is still sufficiently cold that one can observe a sharp Fermi surface in momentum \cite{PhysRevA.86.031601}. Any increase in quasiparticle widths away from $k_F$ will have little effect on the data as long as the quasiparticles have an energy width less than our energy resolution, which should be the case for long-lived quasiparticles. Finally, using a quadratic dispersion over a large range of $k$ is supported by the data \cite{PRL_SOM}. 

\begin{figure}
	\includegraphics[scale=0.7]{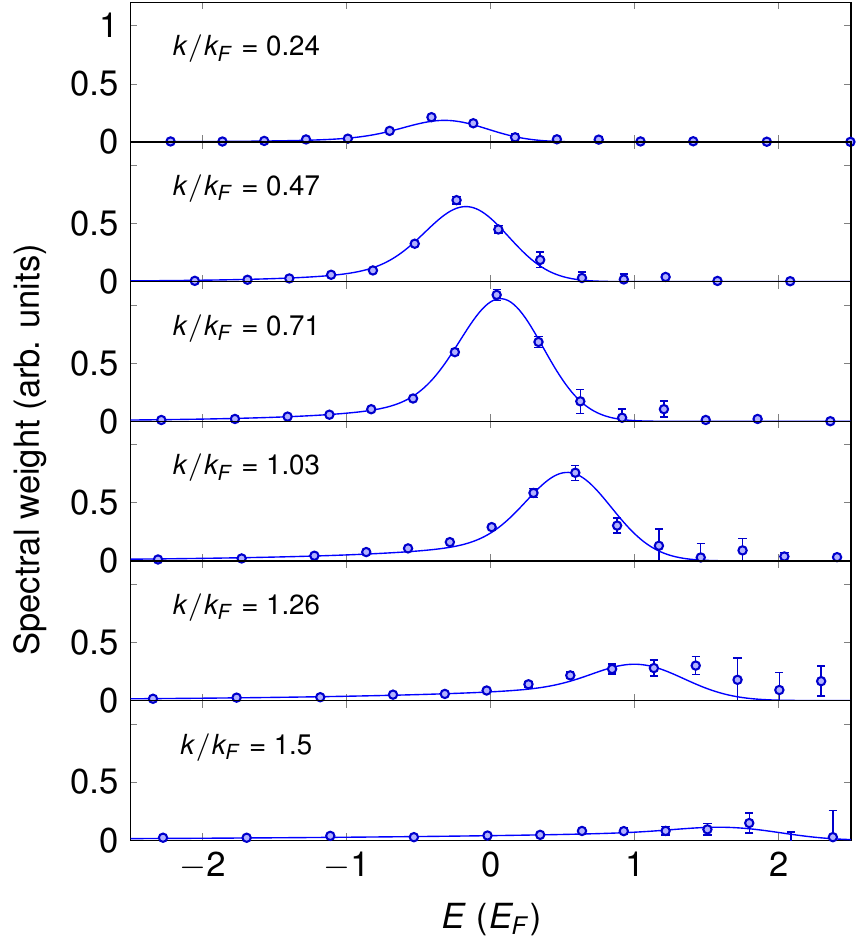}
    \caption{\textbf{EDCs for atom PES data near unitarity.} Data (circles) and fits (lines) are shown for several example traces at fixed $k$ through the 
    PES data at $(k_Fa)^{-1}=-0.08$. These traces are often called energy distribution curves, or EDCs.
     Here, the fitting parameters are $Z=0.37\pm0.03$, $m^*=1.22\pm0.03$, $T=0.24\pm0.02$, $E_0=-0.33\pm0.02$, $\mu=0.19\pm0.04$, $E_p=0.23\pm0.04$, $T_p=1.09\pm0.08$, where the error margins are for one standard deviation and also include a $5\%$ uncertainty in $E_F$. For this fit, the reduced $\chi^2$ is 1.2.}
    \label{figure2}
\end{figure}

The second part in Eq. \ref{fitting_function} needs to accommodate the remainder of the signal, which is often referred to as an ``incoherent background" in a Fermi liquid description. For fermions with contact interactions, one expects an incoherent background at high momentum
due to short-range pair correlations \cite{Tan08,Braaten2012,PhysRevA.81.021601}. Motivated by this and by the normal state in the BEC limit, we use for $I_{\rm{incoherent}}$ a function that describes a thermal gas of pairs. The pairs have a wave function that decays as $\exp\left(-r\right/R)$, where $r$ is the relative distance and $R$ is the pair size \cite{PhysRevA.71.012713}, and a Gaussian distribution of center-of-mass kinetic energies characterized by an effective temperature $T_p$. This gives
\begin{widetext}
\begin{equation}
I_{\rm{incoherent}}(k,E)=\Theta\left(-E_p-E+k^2\right)\frac{8 k \sqrt{\frac{E_p}{T_p}} 
   \exp(\frac{E_p+E-3 k^2}{T_p}) \sinh \left(\frac{2
   \sqrt{2} k
   \sqrt{-E_p-E+k^2}}{T_p}\right)}{\pi ^{3/2}
   \left(E-k^2\right)^2}\ \ ,
\end{equation} 
\end{widetext}
where $\Theta$ is the Heaviside step function, $E_p$ is a pairing energy defined by $k_F R=\sqrt{2/E_p}$, and both $E_p$ and $T_p$ are dimensionless fitting parameters (see Supplementary Material~\cite{PRL_SOM}). 
While this description of the incoherent piece may not fully capture the microscopic behavior except in the BEC limit, we find nonetheless that Eq. \ref{fitting_function}, after convolution with a Gaussian function that accounts for our energy resolution, fits the data very well throughout the crossover.
For each value of $(k_F a)^{-1}$, we perform a surface fit to the roughly 300 points that comprise the PES data $I(k,E)$ for $k\leq1.5$ and $E\geq-3$. The reduced chi-squared statistic, $\chi^2$, after accounting for the seven fit parameters, is between 0.75 and 1.3. 
An example fit is shown in Fig. 2, where we show several traces at fixed $k$ for PES data taken near unitarity.

In Fig. 3a, we show $Z$ as a function of $(k_Fa)^{-1}$. For our lowest $(k_Fa)^{-1}$, $Z\approx0.8$; however, $Z$ decreases rapidly going from the BCS side of the crossover (negative $a$) to the BEC side (positive $a$), reaching $Z \approx 0.3$ at unitarity.
Beyond $(k_Fa)^{-1}=0.28\pm0.02$, $Z$ vanishes, signalling the breakdown of a Fermi liquid description. 
Restricting the fitting to a smaller region around $k_F$ gives results for $Z$ that are consistent with the fits to $k\leq1.5$ (see Fig. 3a). We note that the interaction strength where $Z$ vanishes, as well as the sharpness with which $Z$ goes to zero, are likely to be temperature dependent \cite{arXiv1411.7207}.
The best fit values for the effective mass, $m^*$ are shown in Fig. 3b, where $m^*$ increases with increasing interaction strength as expected for a Fermi liquid. A linear fit gives $m^*=1.21\pm0.03$ at unitarity, which is somewhat higher than $m^*=1.13\pm0.03$ measured in Ref.~\cite{ISI:000275108400031}, but close to the $T=0$ prediction of $m^*=1.19$ from Ref.~\cite{PhysRevA.80.063612}. The other fit parameters for the two-mode function are shown in Fig. S4 in Ref.~\cite{PRL_SOM}.

\begin{figure}
	\includegraphics[scale=1.2]{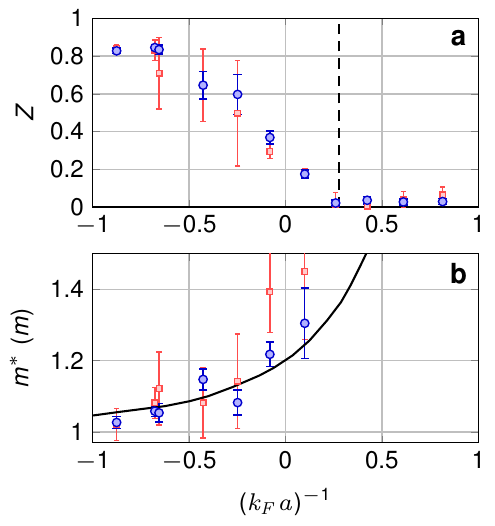}
	\caption{\textbf{Z and effective mass.
		} \textbf{a}, The quasiparticle spectral weight $Z$ decreases as $(k_Fa)^{-1}$ increases. 
		Using a linear fit to the range $-0.5\le(k_Fa)^{-1}\le0.3$, we find that $Z$ vanishes at $(k_Fa)^{-1}=0.28\pm0.02$ (dashed line); this marks the breakdown of a Fermi liquid description. 
		The blue circles come from fits of a large range of data, from 0 to 1.5 $k_F$. The red squares show the result of restricting the fit to 0.7 to 1.3 $k_F$, and they show a similar trend and slightly larger error bars.
		\textbf{b}, The quasiparticle effective mass $m^*$ is shown for the region where $Z>0$. Interactions increase $m^*$, and the data (circles) agree surprisingly well with the theoretical prediction for the limiting case of the Fermi polaron (solid line) \cite{0295-5075-88-6-60007,Navon07052010}. Restricting the fit to EDCs close to the Fermi surface produces similar results with increased error bars (red squares).}
	\label{figure3}
\end{figure}

We note an interesting comparison of our results with the Fermi polaron, which is the quasiparticle in the limit of a highly imbalanced Fermi gas. Schirotzek \textit{et al.} measured $Z=0.39\pm0.09$ for the Fermi polaron at unitarity \cite{PhysRevLett.102.230402}, which is similar to our result for the balanced Fermi gas. For the polaron case, $Z$ also goes to zero in a similar fashion to our results, but
farther on the BEC side of the crossover \cite{PhysRevLett.102.230402}. This similarity is surprising because we expect a phase transition from polarons to molecules in the extreme imbalance limit \cite{PhysRevB.77.020408,PhysRevB.77.125101}, with $Z$ acting as an order parameter \cite{PhysRevA.80.053605}, while, in contrast, the balanced Fermi gas should exhibit a continuous crossover. For $m^*$, we also find that our result is close to the measured effective mass of the Fermi polaron at unitarity \cite{ISI:000275108400031}, $m^*=1.20\pm0.02$, and similar to the predicted polaron mass \cite{0295-5075-88-6-60007,Navon07052010} throughout our measurement range (solid line in Fig. 3b).

\begin{figure}
	\includegraphics[scale=1.4]{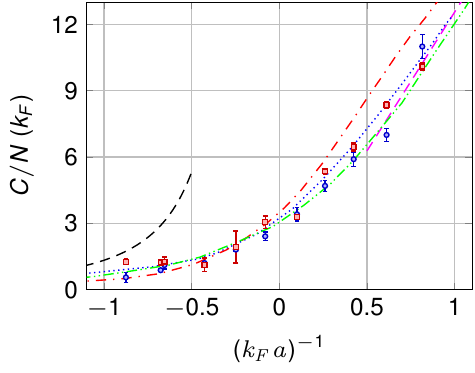}
	\caption{  \textbf{The contact parameter.} From the PES data, we extract the contact per particle (in units of $k_F$) 
		for a homogeneous Fermi gas above $T_c$ 
		as a function of the interaction strength $(k_Fa)^{-1}$. 
		The contact measured from the tail of the rf lineshape \cite{PRL_SOM},
		using data for $h\nu\ge5E_F$, is shown in blue circles, while the contact extrapolated from the fits of the PES data is shown in red squares. Remarkably, even though we limited the fits of the PES data to $k\leq1.5$ and $E\geq-3$, a region with a relatively small contribution of short-range correlations \cite{Tan08,Braaten2012,PhysRevLett.104.235301,PhysRevLett.109.220402}, we find that the contact from the PES fits is consistent with the contact measured from the tail of the momentum-integrated rf line shape. For comparison with the data, we also plot the BCS (dashed black line) and BEC (dashed magenta line) limits, given by $ 4(k_Fa)^2/3$ and $4\pi (k_F a)^{-1}$, respectively \cite{Braaten2012}, the non-self-consistent t-matrix at $T=0$ (dotted blue line) and its Popov version at $T_c$ (dash-dotted red line) \cite{PhysRevA.82.021605-2010}, and the self-consistent t-matrix model at $T=0$ (double-dotted green line) \cite{PhysRevA.80.063612}. Interestingly, we find that the contact measured above $T_c$ agrees well with the $T=0$ theories.}
	\label{figure4}
\end{figure}

As $(k_Fa)^{-1}$ increases, short-range correlations are expected to increase. This gives rise to increased weight in the high-$k$ part of the spectral function \cite{PhysRevA.81.021601}, which is quantified by a parameter called the contact \cite{Tan08,Braaten2012,PhysRevLett.104.235301,PhysRevLett.109.220402}. In a Fermi liquid description, the contact must be accounted for by the incoherent part of the spectral function \cite{PhysRevA.81.021601}. 
We note that our particular choice for $I_{\rm{incoherent}}$ has the expected form of a
 $1/k^4$ high-$k$ tail in the momentum distribution \cite{Tan08} and a $1/\nu^{3/2}$ large-$\nu$ tail in the rf line shape \cite{Braaten2012}, where $\nu$ is the rf detuning. 
 Remarkably, we find that the contact can be accurately extracted from the fits to the PES data even though we restrict the fits to $k\leq1.5$. For comparison, $1/k^4$ behavior in the momentum distribution was observed for $k > 1.5k_F$ \cite{PhysRevLett.104.235301}.
In Fig. 4, we plot the measured contact per particle, $C/N$, in units of $k_F$, as a function of $(k_Fa)^{-1}$. The data extend previous measurements of the contact at unitarity \cite{PhysRevLett.109.220402,PhysRevLett.106.170402} and agree well with several theoretical predictions \cite{PhysRevA.82.021605-2010,PhysRevA.80.063612}.

The results presented here can explain how different observations lead to different conclusions regarding the nature of the normal state of the unitary Fermi gas. Although the data here taken just above $T_c$ show that a Fermi liquid description breaks down for $(k_Fa)^{-1} \ge 0.28\pm0.02$, $Z$ remains finite at unitary. Fermionic quasiparticles may play a key role in thermodynamics, while PES data reveal back-bending and significant spectral weight in an ``incoherent" part that is consistent with pairing. 
With the nearly homogeneous PES data, we find that $Z$ vanishes surprisingly abruptly and note some similarity to Fermi polaron measurements.
Comparing the PES data with various BCS-BEC crossover theories may help elucidate these observations and advance quantitative understanding of the crossover. 

\section*{Acknowledgements}
This work was supported by the National Science Foundation under Grant Number 1125844 and by the National Institute of Standards and Technology.

\bibliography{mybibfile}

\clearpage

\begin{widetext}
	

\section*{Supplementary material}

# Supplementary material for "Breakdown of Fermi liquid description for strongly interacting fermions"

Yoav Sagi, Tara E. Drake, Rabin Paudel, Roman Chapurin, and Deborah S. Jin

*JILA, National Institute of Standards and Technology and the University of Colorado, and the Department of Physics, University of Colorado, Boulder, CO 80309-0440, USA*

# SUPPLEMENTARY METHODS

The experiments are performed with an ultracold gas of $^{40}$K atoms in an optical dipole trap. The trapping frequencies are 243 Hz and 21.3 Hz in the radial and axial directions, respectively, with the axial direction of the trap oriented horizontally. The final stage of evaporative cooling is performed at a magnetic field of 203.3 G, after which the field is swept linearly in 50 ms to the final value where we carry out the PES measurement. The initial temperature of the weakly interacting gas, before the adiabatic sweep, is $T_0 = (0.16\pm0.02)T_F$ [an exception is the farthest point on the BCS side, where $T_0 = (0.13 \pm 0.02)T_F$]. From our previous measurements of the release energy at unitarity [21], we estimate that the temperature at unitarity is $(0.18 \pm 0.02)T_F$. The gas has 80,000 to 120,000 atoms per spin state and is cooled such that the temperature is just above $T_c$ after a sweep to the Fano-Feshbach resonance at 202.20 G [12]. We verified that for all interaction strengths the condensate fraction [12,24] is less than 1%. Additional details on our experimental setup can be found in Refs. [12,13,25].

The atom PES measurement uses an rf pulse with a Gaussian field envelope that has an rms width of 51 $\mu$s (17 $\mu$s), and a total duration of 300 $\mu$s (100 $\mu$s), for $\nu < 34$ kHz ($\nu \geq 34$ kHz) (see Fig. S1). We use shorter pulses at higher $\nu$ to minimize effects due to the motion of the spin-flipped atoms during the rf pulse. The rf detuning $\nu$ is given with respect to the resonant transition frequency $\nu_0$, which is measured for a spin polarized gas. For each PES measurement, we take data for the same 25 values of $\nu$, each of which is repeated three times. The rf power is increased for larger $\nu$, while keeping the fraction of spin-flipped atoms less than 40%. As in Ref. [21], we measure the dependence of the number of spin-flipped atoms on the rf power, and scale the measurements done at different rf powers to correspond to a common level.

Immediately after the rf pulse, and exactly at trap release, we pulse on the hollow-core light beams, which propagate perpendicularly to each other and intersect at the center of the cloud [20,21]. The frequency of these beams is chosen to optically pump atoms from the $|9/2, -5/2\rangle$ state into the upper hyperfine manifold ($F = 7/2$), where they are invisible to our imaging (see Fig. S1). Both the rf pulse and the spatially selective optical pumping, which is pulsed for 40 $\mu$s, are completed in a timescale that is short compared to motion of atoms in the trap. We typically probe the $|9/2, -5/2\rangle$ atoms that came from the central

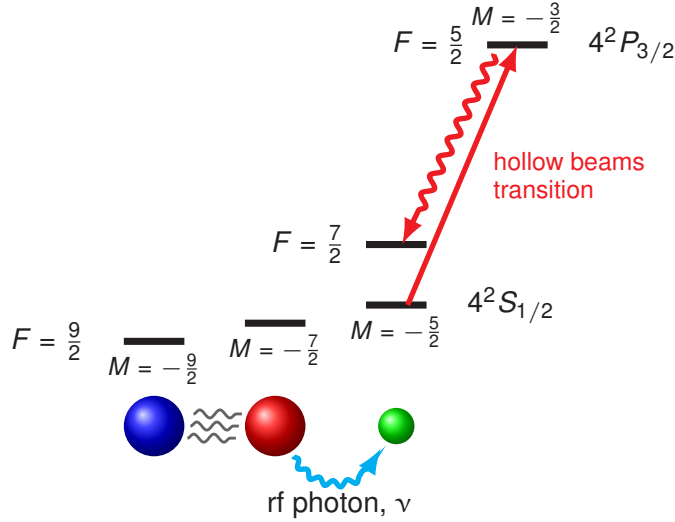

FIG. S1. **Schematic level diagram of $^{40}$K energy states relevant to the experiment.** The gas is prepared in an equal mixture of two hyperfine states. These two states, that can be made strongly interacting near a Feshbach resonance, are $|9/2, -9/2\rangle$ and $|9/2, -7/2\rangle$, where the first number is the total atomic spin $F$ and the second number is its projection $M$. For photoemission spectroscopy, an rf pulse drives the transition from the $|9/2, -7/2\rangle$ state to the $|9/2, -5/2\rangle$ state, which is weakly interacting with the two other spin states. Also shown are the transition driven by the hollow-core light beams (straight red arrow) and the spontaneous emission transition with the largest Clebsch-Gordan coefficient (curvy red arrow).

30% of the cloud; we find this fraction to be a good compromise between spatial selectivity and signal-to-noise ratio [20,21]. As described in Ref. [21], just before imaging the cloud, we remove the remaining atoms from the $|9/2, -9/2\rangle$ and $|9/2, -7/2\rangle$ states and then transfer the outcoupled atoms in the $|9/2, -5/2\rangle$ state to the $|9/2, -9/2\rangle$ state with two short rf $\pi$-pulses. This procedure enables us to image the atoms on the cycling transition, which improves the signal-to-noise ratio.

We report quantities relative to $E_F$ and $k_F$, which are calculated from the average density of the probed gas. Similar to Ref. [21], the average density is determined from the central part of the *in-situ* density distribution of the gas, which is found by measuring the cloud after expansion. In the BCS-BEC crossover, the expansion is hydrodynamic and we deduce the *in-situ* density distribution by solving the Euler equation with a scaling solution that assumes the chemical potential is proportional to $n^\gamma$, where $n$ is the density, and $\gamma$ depends

on the interaction strength [5]. For the purpose of this work, we use $\gamma$ approximated by the Leggett ansatz [5].

## COMPARISON TO TRAP-AVERAGED PES DATA

To compare the nearly homogeneous data with previously published trap-averaged data, we use the analysis described in Ref. [12]. Specifically, independent gaussian fits to fixed-$k$ traces through the PES data are used to find a dispersion $E(k)$. Fig. S2a shows the nearly homogeneous PES data for $(k_Fa)^{-1} = 0.1$, with white points marking the centers from the gaussian fits. Fig. S2b shows the fixed-$k$ traces through the data, which are often called energy distribution curves, or EDCs. These plots can be compared to figures 1 and 2 of Ref. [12], which show trap-averaged data at $(k_Fa)^{-1} \approx 0.15$ and $T/T_C = 1.24$. We find that the PES results for a nearly homogeneous gas are qualitatively similar to the trap-averaged results. Specifically, in both cases we see dispersions that exhibit large energy widths and back-bending.

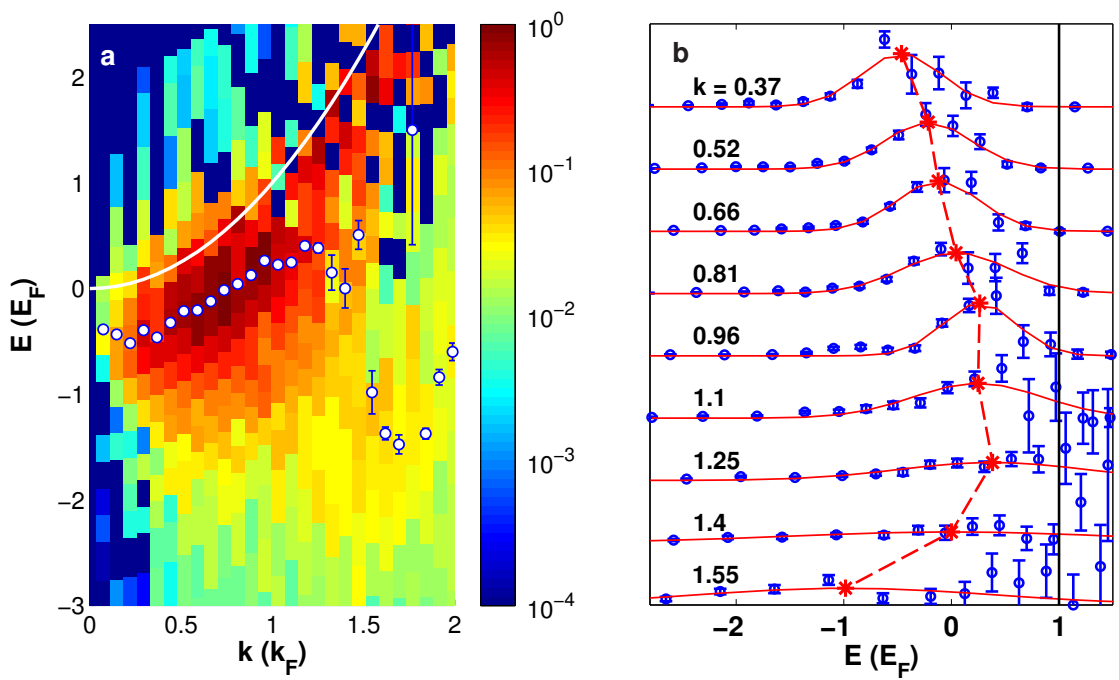

FIG. S2. **Gaussian fits to PES data at $(k_Fa)^{-1} = 0.1$. a**, The white circles indicate the centers from weighted gaussian fits to EDCs at fixed $k$. The white line shows the free-particle dispersion, $E = k^2$. **b**, Individual EDCs (blue points) are shown, along with the fitted gaussians (red lines). Here, each EDC is individually normalized to have the same area, as in Ref. [12]. A solid black line marks $E = E_F$. Red stars show the center of each gaussian, and the dashed red line is a guide to the eye.

5

In the two-mode model, the combination of the positively dispersing coherent part and the incoherent part can produce a spectral function that exhibits back-bending. To demonstrate this, in Fig. S3 we calculate the dispersion with the two-mode model using parameters similar to those from our fits and find the peak using the same analysis technique as was in Ref. [12]. The resulting dispersion clearly shows a BCS-like back-bending.

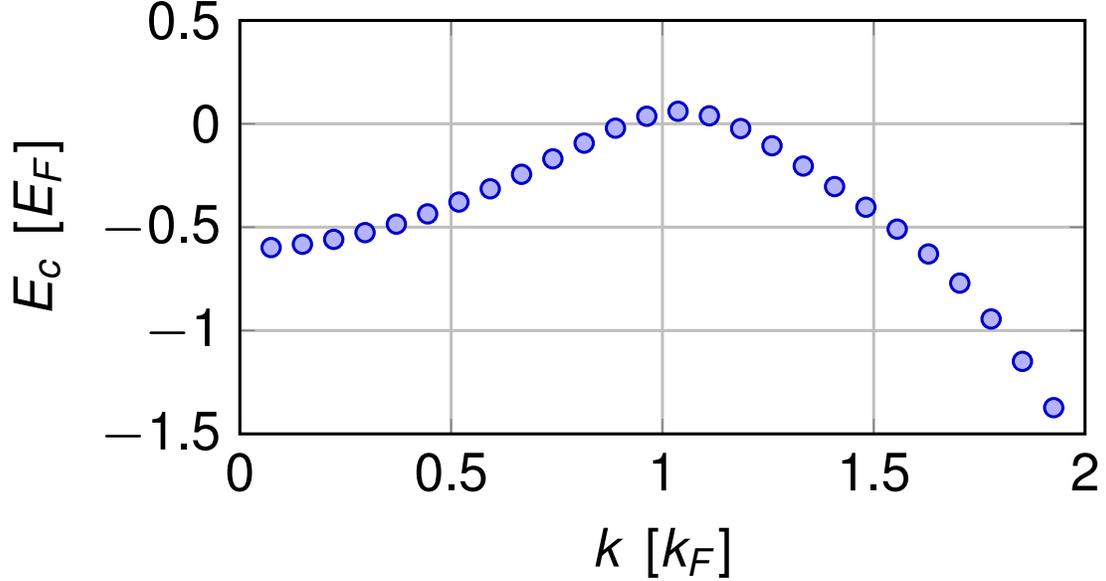

FIG. S3. **Back-bending in the two-mode model.** We generate a simulated spectral function using the two-mode model with the following parameters: $Z = 0.2$, $m^* = 1.2$, $T = 0.25$, $E_0 = -0.5$, $\mu = -0.5$, $E_p = 0.5$, $T_p = 0.75$. Each point in this figure was then obtained by fitting the spectral function at a given momentum $k$ by the Gaussian function $Ae^{-(E-E_c)^2/2\sigma^2}$. In the fits, the value of $\sigma$ was constrained to be above the experimental resolution.

**QUADRATIC DISPERSION**

We fit the PES data to a two-mode function that assumes a quadratic dispersion of the Fermi liquid quasiparticles. While this fit agrees well with the data, we can alternatively locate the peak positions in the EDCs at each $k$ to map out the dispersion $E(k)$, without assuming that it is quadratic. To focus on a quasiparticle-like peak, and to reduce the effect of the asymmetric tails, we use a Gaussian fit where the width is constrained to be equal to our resolution ($h \times 2400$ Hz) and where we also include a constant offset as a fit parameter. An example of the resulting peak dispersion for $(k_F a)^{-1} = -0.43$ is shown in Fig. S4. We can then fit the peak dispersion to $A k^\alpha + B$ to extract an exponent. The results are shown in the inset of Fig. S4 as a function of $(k_F a)^{-1}$, where we restrict this analysis to the region where $Z > 0$. (For higher $(k_F a)^{-1}$, the EDCs are increasingly asymmetric and the peak positions from the gaussian fits are noisy.) We find that $\alpha$ is consistent with 2, which matches the quadratic dispersion of Fermi liquid theory. Interestingly, Fermi liquid theory is typically only applied in the region near $k = k_F$, while the peak dispersion is quadratic.

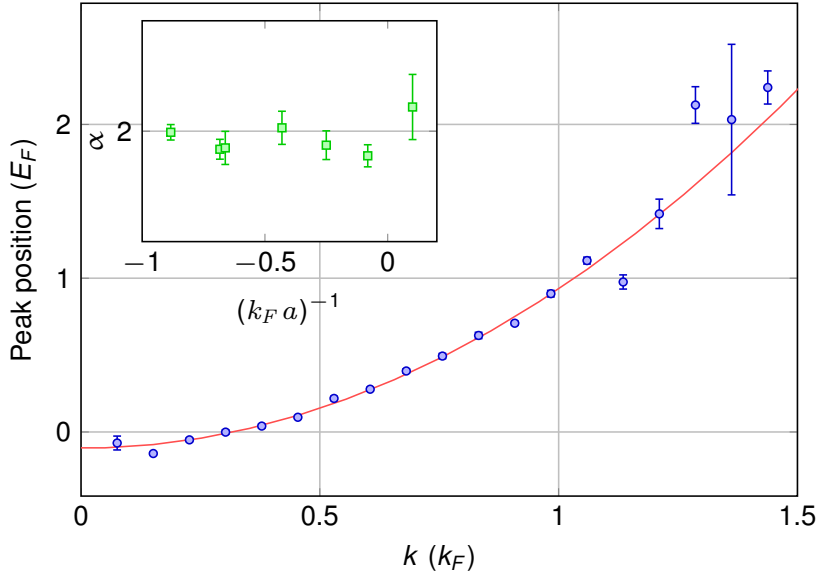

FIG. S4. **Observation of quadratic dispersion.** An example for the peak dispersion, which is determined by finding the peak location for each $k$, is shown for $(k_F a)^{-1} = -0.43$ (circles). We locate the peaks by fitting each EDC separately to a Gaussian plus an offset, where the Gaussian width is fixed to the experimental resolution. The red line shows a fit to $Ak^\alpha + B$ in the range $0.2 < k < 1.5$. **Inset,** The exponent from fits to the peak dispersion (squares). The error bars indicate $\pm$ 1 standard deviation. We find that $\alpha$ is consistent with 2.

# DERIVATION OF THE SINGLE-PARTICLE SPECTRAL FUNCTION FOR PAIRS

For the incoherent component in our two-mode fits to the PES data, we use a model of the PES signal for pairs. To derive this model, we first consider a single stationary pair that is dissociated by absorbing an rf photon. For rf dissociation, energy conservation gives

$$-E_p + h\nu = E_{\text{rel}}, \tag{1}$$

where $E_{\text{rel}} = \frac{\hbar^2 k_{\text{rel}}^2}{m}$ is the relative kinetic energy of the two resulting atoms after dissociation and $E_p = \frac{\hbar^2}{mR^2}$ is the binding energy corresponding to a bound wave function $\phi(r) = \sqrt{\frac{2}{R}} e^{-r/R}$. On the other hand, in PES [13], the single-particle energy $E$ is given by

$$E + h\nu = \frac{\hbar^2 k^2}{2m}, \tag{2}$$

where $\hbar k$ is the momentum of the spin-flipped atom. If the pair is initially at rest, the two resulting atoms will have momenta $\hbar \mathbf{k}_{\text{rel}}$ and $-\hbar \mathbf{k}_{\text{rel}}$, so that combining Eqs. 2 and 1, we find $E = -E_p - \frac{\hbar^2 k^2}{2m}$. Here, we can see that pair dissociation yields a negative, quadratic single-particle dispersion.

The amplitude of the signal as a function of $k$ can be obtained from the normalized rf line shape for the dissociation of a weakly bound molecule, which was derived by Chin and Julienne [35]:

$$F(E_{\text{rel}}) = \frac{2}{\pi} \frac{\sqrt{E_{\text{rel}} E_p}}{(E_{\text{rel}} + E_p)^2}, \tag{3}$$

where $\int_0^\infty F(E_{\text{rel}}) dE_{\text{rel}} = 1$ and we have assumed that the final state is weakly interacting. Eq. 3 can be rewritten in terms of $k_{\text{rel}}$ to give

$$G(k_{\text{rel}}) = \frac{\hbar^3}{\pi^2 m^{3/2}} \frac{\sqrt{E_p}}{(\hbar^2 k_{\text{rel}}^2/m + E_p)^2}, \tag{4}$$

where $\int_0^\infty G(k_{\text{rel}}) 4\pi k_{\text{rel}}^2 dk_{\text{rel}} = 1$. This lineshape is highly asymmetric and has an extent in $k_{\text{rel}}$ that scales as $\sqrt{E_p}$, or equivalently $1/R$. At high $k_{\text{rel}}$, $G(k_{\text{rel}})$ falls off as $1/k_{\text{rel}}^4$, which is consistent with Tan's contact.

If the pair is moving, then combining Eqs. 2 and 1 gives

$$E = -E_p - \frac{\hbar^2 k_{\text{rel}}^2}{m} + \frac{\hbar^2 k^2}{2m}. \tag{5}$$

In PES, we apply rf with detuning $\nu$, measure $k$ for the spin-flipped atom, and extract $E$ using Eq. 2. From Eq. 5, we see that for the dissociation of non-stationary pairs, data with a particular $k$ (but variable $\nu$) can, in principle, yield $E$ anywhere in the range $-E_p - \frac{\hbar^2 k^2}{2m} \leq E \leq -E_p + \frac{\hbar^2 k^2}{2m}$, where the lower limit corresponds to a stationary pair and the upper limit corresponds to a pair with the maximum possible center-of-mass momentum, $\hbar \mathbf{K}_{\mathrm{cm}} = 2\hbar \mathbf{k}$.

To model the center-of-mass momentum distribution of non-condensed pairs, we assume a Maxwell-Boltzmann distribution with an effective temperature $T_p$:

$$P(K_{\mathrm{cm}}) = \left(\frac{\hbar^2}{4\pi k_b m T_p}\right)^{3/2} e^{-\frac{\hbar^2 K_{\mathrm{cm}}^2}{4 k_b m T_p}}, \tag{6}$$

where $k_b$ is the Boltzmann constant and $\int_0^\infty P(K_{\mathrm{rel}}) 4\pi K_{\mathrm{cm}}^2 \mathrm{d}K_{\mathrm{rel}} = 1$. The product of the functions in Eqs. 6 and 4 gives the transition probability as a function of $\mathbf{K}_{\mathrm{cm}}$ and $\mathbf{k}_{\mathrm{rel}}$. We change variables to $\mathbf{k}$ and $\mathbf{k}_{\mathrm{rel}}$ and integrate over the angles to find the transition probability as a function of $k$ and $k_{\mathrm{rel}}$:

$$H(k, k_{\mathrm{rel}}) = \frac{4\hbar^4}{\pi^{3/2} m^2} \sqrt{\frac{E_b}{k_b T_p}} \frac{k k_{\mathrm{rel}} e^{-\frac{2\hbar^2 (k^2 + k_{\mathrm{rel}}^2)}{k_b m T_p}} \left( e^{\frac{\hbar^2 (k+k_{\mathrm{rel}})^2}{k_b m T_p}} - e^{\frac{\hbar^2 (-k+k_{\mathrm{rel}})^2}{k_b m T_p}} \right)}{(\hbar^2 k_{\mathrm{rel}}^2/m + E_b)^2}, \tag{7}$$

where $\int_0^\infty \int_0^\infty H(k, k_{\mathrm{rel}}) \mathrm{d}k \mathrm{d}k_{\mathrm{rel}} = 1$. Finally, we change variables to $E$ and $k$ to derive the PES signal (Eq. 3 in the main text):

$$I_p(k, E) = \Theta\left(-E_p - E + k^2\right) \frac{8k \sqrt{\frac{E_p}{T_p}} e^{\frac{E_p + E - 3k^2}{T_p}} \sinh\left(\frac{2\sqrt{2} k \sqrt{-E_p - E + k^2}}{T_p}\right)}{\pi^{3/2} (E - k^2)^2}. \tag{8}$$

In this last step, we also changed to dimensionless parameters ($k$, $E$, $E_p$, and $T_p$) normalized by $k_F$ and $E_F$.

**FITS TO THE PES DATA**

The main text presents two of the seven fit parameters used to describe our PES data. In this section, we discuss the remaining parameters, which are shown as a function of $(k_F a)^{-1}$ in Fig. S5.

We start with the two temperatures $T$ and $T_p$, which are shown in Fig. S5**a**. For the quasiparticle component, $T$ is close to, but somewhat above the estimated temperature of the gas, $T = (0.18 \pm 0.02)T_F$. We attribute the discrepancy to the remaining density inhomogeneity of the probed portion of the trapped gas, which can broaden sharp features. In particular, when applying our PES measurement technique to a weakly interacting gas, we find that the resulting momentum distribution has a step at the Fermi surface whose width typically corresponds to $T \approx 0.25 T_F$. For the incoherent component, $T_p$, which in the model characterizes the spread of center-of-mass energies of pairs, is around $0.75 T_F$. This is much higher than the temperature of the gas and suggests that the data cannot be simply interpreted as coming from weakly bound bosonic molecules in thermal equilibrium with a Fermi gas of atoms.

Fig. S5**b** shows the second fit parameter in $I_{\text{incoherent}}$, which is a binding energy $E_p = \frac{\hbar^2}{mR^2}$ that sets the size of the pairs $R$. For the data farthest on the BEC side of the crossover, $E_p$ approaches the two-body molecule binding energy, $\frac{\hbar^2}{ma^2}$ (solid line in Fig. S5**b**). In the crossover, $E_p$ is larger than the binding energy for two-body molecules, as one would expect for many-body pairs.

The two remaining fit parameters, $E_0$ and $\mu$, come from the quasiparticle part of the fitting function and are shown in Fig. S5**c** and **d**, respectively. The Hartree energy shift $E_0$ is negative and has a magnitude that increases as $(k_F a)^{-1}$ increases, reaching a value of $-0.46 \pm 0.03 E_F$ at unitarity, in good agreement with several theoretical calculations [18,26-28]. Fig. S5**c** shows $E_0$ for the full range of $(k_F a)^{-1}$ for our data. Here, one can clearly see the jump in the fit values and the large uncertainties that appear in the region where the fits return a small $Z$. This happens because the quasiparticle component that is fit for $(k_F a)^{-1} < 0.2$ has vanished. For this reason, when discussing the fit parameters for the quasiparticle part ($m^*, T, E_0$, and $\mu$), we focus on the region $(k_F a)^{-1} < 0.2$.

The chemical potential, $\mu$, from the quasiparticle part of the fit is shown in Fig. S5**d**. In general, one expects $\mu$ to decrease as $(k_F a)^{-1}$ increases. The data (blue circles) for

11

$(k_F a)^{-1} > -0.5$ follow this expected trend, but the points for our two lowest $(k_F a)^{-1}$ values are surprisingly low. In addition, all the values of $\mu$ here are somewhat lower than the predicted chemical potential in the BCS-BEC crossover [7]. We have found that there is some interdependence of the two fit parameters $\mu$ and $T$ for the quasiparticles. This is illustrated in Fig. S5**d**, where the red squares show the result for $\mu$ when we fix $T = 0.25$ in the fits. (The effect of fixing $T$ on the other fit parameters is to reduce some scatter but is otherwise minimal.) This interdependence, and the fact that $T$ is increased by technical issues (namely, the remaining density inhomogeneity that limits our resolution in $k$), may play a role in explaining the lower than expected values for $\mu$.

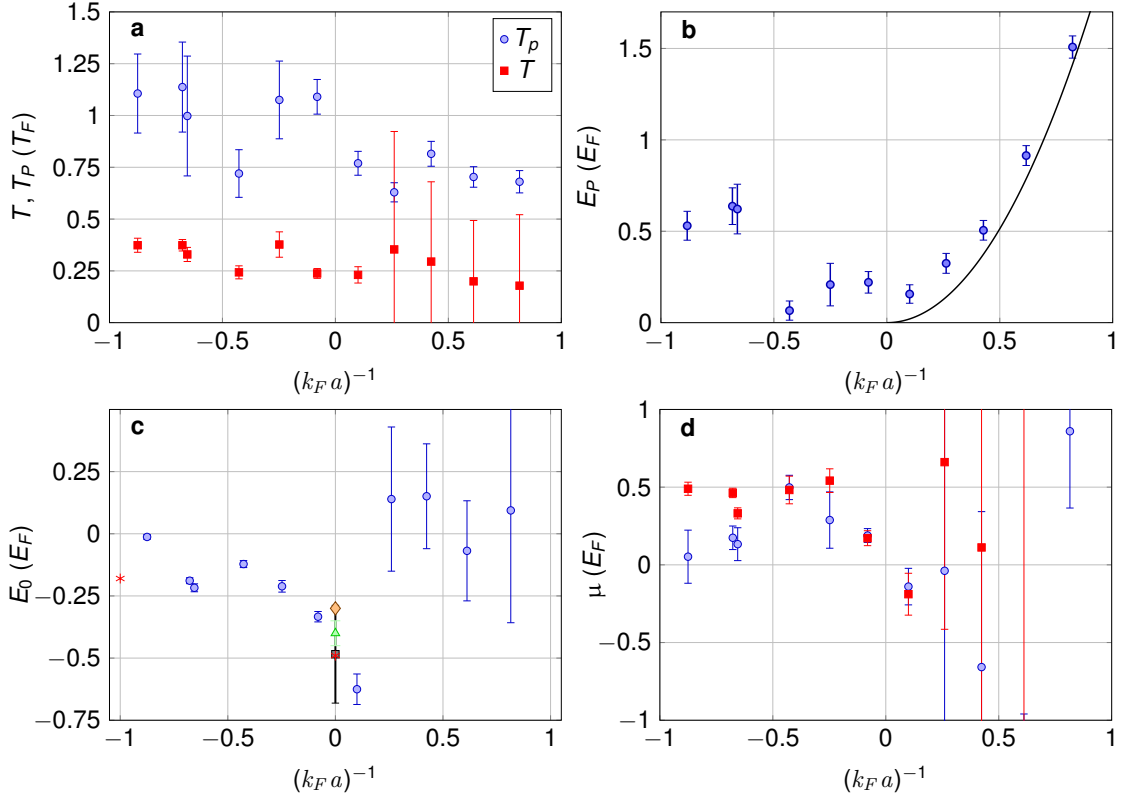

FIG. S5. **Parameters of the fits to the PES data. a**, We extract two temperatures from our fits to $I(k,E)$. The quasiparticle parts of the fits to the PES data give $T$ values (red squares) that are comparable to, but higher than, our estimated temperature of the gas at unitarity ($T = 0.18 \pm 0.02$). In contrast, the incoherent parts of the fit give $T_p$ values (blue circles) around 0.75. $T_p$ can be thought of as an effective temperature that characterizes the energy width of the incoherent part of the spectral function. **b**, The pairing energy, $E_p$, comes from the incoherent part of the fit. At the largest $(k_F a)^{-1}$, the data (circles) approach the two-body molecular binding energy (line). **c**, The Hartree energy shift, $\mathbf{E_0}$, comes from the quasiparticle part (blue circles). The error bars correspond to $\pm$ 1 standard deviation from the fits and 5% uncertainty in $E_F$. Also shown are the results of a quantum Monte-Carlo calculation [26] (black square) and three theories by Haussmann *et al.* [18] (red asterisks), Kinnunen [27] (green triangle), and Bruun and Baym [28] (orange diamond). **d**, The chemical potential $\mu$ is also found in the quasiparticle part of the fit. We show the results of the fit with (red squares) and without (blue circles) the constraint $T = 0.25$. The fact that constraining $T$ affects the best fit value for $\mu$ reveals an interdependence of $\mu$ and $T$ that we do not observe for the other fit parameters.

13

**THE CONTACT**

Fig. 4 of the main text shows the contact measured from the tail of the rf lineshape, $\Gamma(\nu)$. For this we use the same data as for PES, but simply look at the number of spin-flipped atoms as function of $\nu$, summing over all momenta. Fig. S6 shows an example rf lineshape for the data at $(k_F a)^{-1} = -0.25$. We use the contact extracted from the rf lineshapes when we normalize the PES data so that $\int_0^\infty \int_{-\infty}^\infty I(k, E) \mathrm{d}E \mathrm{d}k = 1$. Since the PES data span a finite range of $E$ and $k$ (typically, $-4 < E < 3$ and $k < 4$), knowing the behavior of $\Gamma(\nu)$ for large $\nu$ allows us to account for signal beyond the range of the data in our normalization of the PES data.

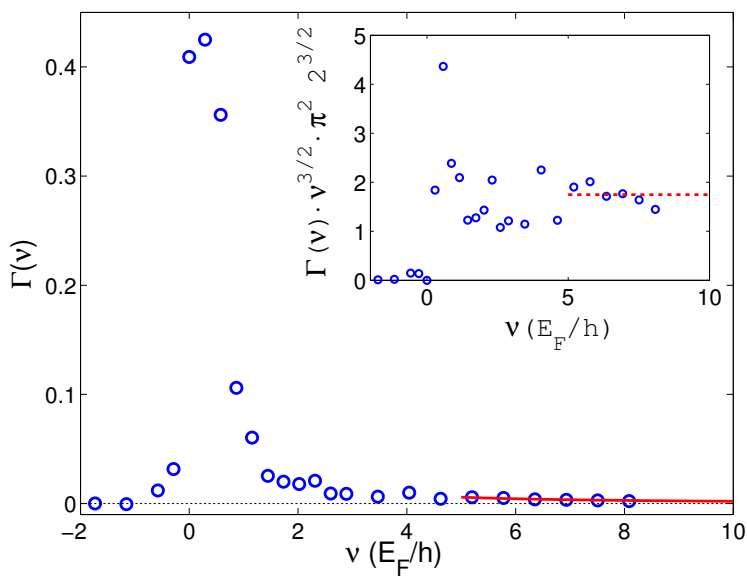

FIG. S6. **The rf lineshape $\Gamma(\nu)$ for $(k_F a)^{-1} = -0.25$**. The contact is extracted by fitting the data for $h\nu \geq 5 E_F/h$ to $(C/N)/(2^{3/2}\pi^2 \nu^{3/2})$, with the normalization $\int \Gamma(\nu) d\nu = 0.5$. The inset shows the signal $\Gamma(\nu)$ multiplied by $2^{3/2}\pi^2 \nu^{3/2}$. The red lines show the fit, and for this dataset, we extract $C/N = (1.81 \pm 0.08) k_F$.

Fig. 4 of the main text also includes the contact extrapolated from the surface fits to the PES data. In the fitting function, the contact can only be accommodated by the "incoherent" component, since the quasiparticle part only includes narrow peaks that are symmetric in energy. For the extrapolation, we use the model, with the best fit parameters for each data set, to calculate the momentum distribution: $n(k) = \int_{-\infty}^\infty I_p(k, E) \mathrm{d}E$ out to $k = 5$. We then find the contact from the average value of $k^4 n(k)$ in the region $3 \leq k \leq 5$. (Note that the

fits were restricted to data for $k < 1.5$.)

As a check of the fitting function, we have compared these results with a simple expression derived from the contact for weakly bound molecules. For a dimer with a wave function $\phi(r) = \sqrt{\frac{2}{R}} e^{-r/R}$, the contact [29] is $C_{\text{dimer}} = 8\pi/R$. Since the dimer consists of two atoms, the contact per particle is $C/N = 4\pi/R$. Expressing this in terms of the dimensionless $E_p$ and multiplying by the amplitude $(1 - Z)$, we have

$$\frac{C}{Nk_F} = 4\pi(1 - Z)\sqrt{\frac{E_p}{2}} \ . \tag{9}$$

We find that the result of Eq. (9) agrees very well with the contact extracted directly from the momentum tail of the fitted dispersion.


\end{widetext}

\end{document}